\documentstyle[11pt,twoside,pasp3D,epsf]{article}
\begin{document}
\title{Critical Comparison of 3-d Imaging Approaches}
\author{Charles L. Bennett}
\affil{L-43, Lawrence Livermore National Laboratory, P.O. Box 808,
Livermore, California, 94550}

\begin{abstract}

Currently three imaging spectrometer architectures, tunable filter,
dispersive, and Fourier transform, are viable for imaging the universe
in three dimensions. There are domains of greatest utility for each of
these architectures. The optimum choice among the various alternative
architectures is dependent on the nature of the desired observations,
the maturity of the relevant technology, and the character of the
backgrounds.  The domain appropriate for each of the alternatives is
delineated; both for instruments having ideal performance as well as
for instrumentation based on currently available technology. The
environment and science objectives for the Next Generation Space
Telescope will be used as a specific representative case to provide a
basis for comparison of the various alternatives.
\end{abstract}

\section{Introduction}

It is expected that within the year, a decision will be made as to the
composition of the suite of science instruments to be deployed on the
Next Generation Space Telescope (NGST). It is therefore a particularly
good time for a discussion of the relative merits, and appropriate
domains of greatest utility for the various 3-d imaging
alternatives. There has been, and no doubt will continue to be, a
great deal of discussion as to which approach to 3-d imaging is ``the
best''. There is no single correct answer, of course, since each type
of instrument has its own strengths and weaknesses.

It does not seem
to be widely known that, in the limiting case of photon statistical
noise dominance, the performance of a 3-d imaging spectrometer based
on 2-d detector arrays is the same for all architectures
(Bennett et al. 1995), whether
tunable filter, dispersive, or Fourier transform, provided that the
same degrees of freedom are measured. In the following, I will first
consider the photon statistics limited case, and show the equivalence
between the various architectures. I will then generalize to the
performance in the case that detector read noise, dark current, and
Zodiacal background are included. I will consider specific parameters
that are appropriate for the anticipated NGST environment. Finally, I
will offer a suggestion for a hybrid instrument which combines the
best features of all of the 3-d architectures, and offers great
potential for best meeting the NGST needs.

\section{Tunable Filter vs. Dispersive Spectrometer (Ideal Limit)}

\begin{figure}
\plotfiddle{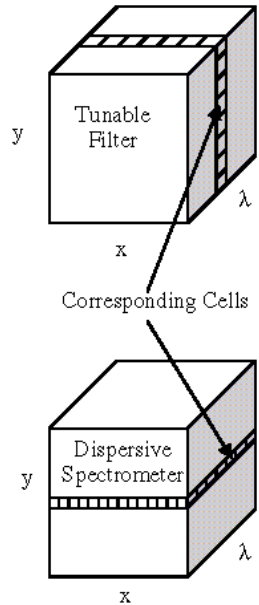}{3.2in}{0}{50}{50}{-80}{-10}
\caption{A schematic comparison between a Tunable Filter 
Imaging Spectrometer and a Dispersive Imaging Spectrometer.
}
\label{Bennett-fig1}
\end{figure}

In comparing between the various options, it is important to assume
equivalent detectors. In order to obtain 3-d data using a 2-d detector
array, a series of exposures must be made. Consider an $N \times M$
pixel focal plane array, having no ``gaps'' between the pixel elements.
Typical frames for a dispersive imaging spectrometer (DS), and a
tunable filter imaging spectrometer (TF) are indicated schematically in
Figure \ref{Bennett-fig1}.  In general, it is of course not necessary
for the spatial samples observed by the DS to be contiguous, as implied
by the arrangement displayed in Figure \ref{Bennett-fig1}.  Nor is it
necessary for the spectral samples observed by the TF to be contiguous
and non-overlapping, as is also implied by the configuration displayed
in Figure \ref{Bennett-fig1}.  Indeed, in some cases, non-contiguous
spectral sampling is desirable, and the TF system lends itself much
more naturally to this mode of operation. On the other hand, for some
questions, the ability to observe non-contiguous spatial samples is
very important, and the DS approach, such as with a Multi-Object
Spectrometer (MOS), is better suited for such measurements. For the
moment, consider the case that the same spatial and spectral samples
are covered by both the TF and the DS. Assume that the spectral samples
represented by the various pixels along the dispersion direction in the
DS correspond exactly both in terms of bandwidth and band center to the
series of measurements made by the TF system, and that the spatial
samples represented by the various pixels in the TF system similarly
correspond exactly to the series of spatial measurements made by the DS
system. In this case, if the total observation time is divided equally
among the spectral samples for the TF case, and for the spatial samples
in the DS case, each cell in the 3-d datacube is observed for the same
exposure time, and with the same efficiency. Clearly the signal to
noise performance will be the same for both of these configurations.

\section{Tunable Filter vs. Fourier Transform Spectrometer (Ideal
Limit)}

The relation between the performance of an ideal tunable filter
spectrometer with an ideal Fourier transform spectrometer is more
subtle than that between the tunable filter and the dispersive
spectrometer. One simplification, however, is that since the size of
the image may be assumed the same for the FT and TF systems, it is
only necessary to consider the information content of a single
representative detector element obtained via either the TF or the FT
system.

It is helpful to
consider an analogy with the use of the Modulation Transfer Function
(MTF) for the characterization of imaging systems. Consider an
``object'' spectrum having a sinusoidal intensity variation as a
function of frequency. Also assume that this object spectrum is
observed with a TF spectrometer having uniformly spaced filter
samples, and that all of the filter samples have an equal transmission
bandwidth. The ``image'' spectrum would also have a sinusoidal intensity
variation as a function of frequency. In the case that the period of
the sinusoidal intensity variation is much smaller than the
characteristic width of the TF spectral channels, the ``image'' spectrum
modulations are greatly reduced. Furthermore, if the spacing of the TF
spectral samples is not sufficiently dense, the period of the
modulations in the ``image'' spectrum may be altered by ``aliasing
effects''.

An FT spectrometer, at each of a sequence of retardance settings,
directly measures the intensity of a particular sinusoidal intensity
variation in the object spectrum. The set of such measurements
constitutes an interferogram. In order to compare the information
content of TF spectra measured in the frequency domain with FT
interferograms measured in the transform domain, it is important to
carefully consider the shape of the spectral response of the TF
filters, their spacing, and the amount of spectral information content
being measured.

In a naive approach to a TF system, it
would be assumed that the transmission function for each of the TF
filters had a ``top hat'' shape, i.e. outside the spectral bandpass of a
given filter the transmission would be zero, and within a given
bandpass the transmission would be unity. Viewed in terms of the
response to sinusoidal modulations in the ``object'' spectrum, such
filters have undesirable ramifications, such as contrast reversal for
some modulation periods, and aliasing for others. Correspondingly, the
most straightforward approach to the acquisition of interferograms by
an FT spectrometer, involving equal weighting of each of the
retardance measurements, produces effective spectral response
functions which have undesirable negative sidelobes. It is important
to consider spectral transmission functions which do not have such
``sharp corners'' as the ``top hat'' shape for the TF case, and to
consider tapered weighting of the interferograms for the FT case.

Consider a sequence of measurements of the intensity of an underlying
continuous spectral intensity function $S(\nu)$, dependent on the
frequency $\nu$, that is transmitted through a spectral filter
$\tau(\nu)$. For a transmission filter centered at $\nu_0$, the
observed number of photoelectrons would be given by

\begin{equation}
S_{\nu_0} = T_{\nu_0} \int_0^\infty S(\nu) \tau(\nu_0 - \nu) d\nu.
\end{equation}
Here the units of spectral radiance $S(\nu)$ are photons Hz$^{-1}$
s$^{-1}$, the exposure time for the observation is $T_{\nu_0}$, in
units of s, while the transmission function $\tau(\nu)$ is
dimensionless.  Also, although the integration limits extend to
infinity, this is a purely formal convenience, and in this integral,
as in others to follow, the integrand will always be limited to a
finite range. The frequency variable, $\nu$, is in units of Hz.  (It
is sometimes convenient to use the wavenumber equivalent of the
frequency, defined by $\nu/c$, and having dimensions of cycles per
cm). It is assumed that the quantum efficiency is unity. The peak
transmission is assumed to be unity, and the effective width of the
transmission filter may be defined by

\begin{equation}
\Delta\nu_{eff} = \int_0^\infty \tau(\nu) d\nu.
\end{equation}
In the case that the spectral radiance function varies slowly over the
interval for which $\tau(\nu)$ is significant, the integral in Eq. (1)
may be approximated by

\begin{equation}
S_{\nu_0} \simeq T_{\nu_0} S(\nu_0) \Delta \nu_{eff}.
\end{equation}
The variance in the observed number of photoelectrons, in the
statistical noise limit is equal to the total number of photoelectrons
detected,

\begin{equation}
var(S_{\nu_0}) = T_{\nu_0} S(\nu_0) \Delta \nu_{eff}.
\end{equation}
Using the relation between the observed counts and the estimate of the
underlying spectral radiance function evaluated at $\nu_0$
of Eq. (3),

\begin{equation}
var(S(\nu_0)) = {S(\nu_0) \over T_{\nu_0} \Delta \nu_{eff}}.
\end{equation}
For comparison with the FT spectrometer case, for which the noise
spectrum $var(S(\nu_0))$ is independent of $\nu_0$, the dwell time
$T_{\nu_0}$ is taken proportional to $S(\nu_0)$.  (This assumed dwell
time variation could of course only be used if the spectrum is known,
and would not be applicable to multiple pixels, if they contain
different spectra. The impact of varying spectral shape on the
comparison between FT and TF spectrometers will be further discussed
below.) The constant of proportionality may be determined by requiring
that the sum over all $\nu_0$ channels yields the total observation
time,

\begin{equation}
T_{\nu_0} = T_{tot}
{ S(\nu_0) \Delta \nu_0  \over \int_0^\infty S(\nu) d\nu }.
\end{equation}
Here the factor $\Delta\nu_0$ is the spacing between the TF spectral
samples. For this integration time sequence the spectral variance
becomes,

\begin{equation}
var(S(\nu_0)) =
{\int_0^\infty S(\nu) d\nu \over  T_{tot} \Delta \nu_0 \Delta \nu_{eff} }.
\end{equation}
With measurements made at the sample spacing $\Delta\nu_0 = \Delta
\nu_{eff}$ this yields

\begin{equation}
var(S(\nu_0)) =
{\int_0^\infty S(\nu) d\nu \over  T_{tot} \Delta \nu_{eff}^2 }.
\end{equation}.

Measurements made at a sample spacing much finer than this produce
little additional information about the continuum function $S(\nu)$,
since the magnitude of $\Delta\nu_{eff}$ sets a practical limit to the
fineness of the resolution recoverable, no matter how fine the sample
spacing.

\subsection{Derivation of the Basic Fourier Transform Relationships}

The intensity of the interference pattern in a dual output port
Michelson interferometer, $I(x)$, is a continuous function of the
optical path difference $x$, i.e., the retardance, between the two
mirrors, related to the continuous spectral intensity detected,
$S(\nu)$, by the integral,

\begin{equation}
I_\pm(x) = {1\over 2} \int_0^\infty S(\nu) (1 \pm cos( 2\pi \nu x/c)) dx.
\end{equation}
The two output ports correspond to the two sign values, with the ``+''
sign corresponding to the output port for which the two interfering
beams are in phase at zero optical path difference (ZPD), and the ``-''
sign corresponding to the output port with out of phase beams at
ZPD. As before, the product $S(\nu)d\nu$
has units of counts per
second. Eq. (9) is valid for a perfectly compensated, perfectly
efficient beam splitter. Real beam splitters have dispersion and are
not perfectly efficient, but these complications are easily dealt with
in practice. It is convenient to form the sum and difference of the
signals from the two output ports of the interferometer. These two
quantities are given by the integrals,

\begin{equation}
I_\Sigma (x) = \int_0^\infty S(\nu)d\nu,
\end{equation}

\noindent
and

\begin{equation}
I_\Delta(x) = \int_0^\infty S(\nu) cos( 2\pi \nu x/c)) dx.
\end{equation}
Note that the summed signal is independent of the optical path
difference $x$, and is simply given by the integrated spectral
intensity. Thus at each retardance setting of the interferometer the
full broad band image is measured. This is because, in the absence of
absorption losses, every photon entering the interferometer goes to
one or the other of the exit ports. The difference signal, at the zero
retardance position also becomes equal to the same full band intensity
integral.

This feature of the summed signal from an FT system suggests that a
desirable hybrid of FT and TF may be obtained by simply having a
tunable filter placed in the optical train of an imaging FT
spectrometer. In this case, the sum of the two output ports of the FT
spectrometer provides the unmodulated full intensity of the light that
has passed through the tunable filter. In addition, higher resolution
spectral imaging may be obtained at the same time. In this hybrid
approach, the summed output will be called the ``panchromatic'' output
of the FT, while the transform of the difference output will be called
the ``spectral'' output of the FT instrument.

In general, it is advantageous to have the dwell time depend on
retardance in order to tailor the effective spectral line shape and
maximize data collection efficiency. This is typically done for radio
astronomy, but is not typically done for laboratory FTIR
spectroscopy. A typical interferogram would consist of a set of $N$
discrete samples of the continuous function $I(x)$, symmetric about the
point $x=0$, each observed with dwell time $T_n$.

\begin{equation}
I_n = T_n I_\Delta(x_n), {\rm\ with\ } x_n = n \delta x,
\ n\ {\rm in\  the\ range\ } [-N/2+1, N/2].
\end{equation}
Discrete Fourier transformation results in periodogram estimates,
$S_k$, at integer multiples $k$ of a fixed frequency spacing $\delta
\nu$, approximately related to the continuous function $S(\nu)$ by

\begin{equation}
S_k \simeq T_0 S(\nu_k) \delta\nu {\rm\ with\ } \nu_k = k \delta\nu, 
\ k \ {\rm in\ the\ range\ } [-N/2+1, N/2].
\end{equation}
The approximate relation between the discrete spectral estimate and
the continuous spectral function is accurate to the extent that the
continuous spectral function varies sufficiently slowly in the
neighborhood of the discrete sample point at $\nu=\nu_k$.  This
condition is similar to that used in writing expression (3) for the TF
case. The spectral sample spacing $\delta \nu$ and the interferogram
sample spacing $\delta x$ are related by $\delta\nu = c/(N\delta
x)$. The $S_k$ values are given by the discrete Fourier transform,

\begin{equation}
S_k = {2 \over N} \sum_{n = -N/2 + 1}^{N/2} I_n exp(-i 2 \pi kn/N).
\end{equation}
The inverse discrete Fourier transform is

\begin{equation}
I_n = {1 \over 2} \sum_{k = -N/2 + 1}^{N/2} S_k exp(i 2 \pi kn/N).
\end{equation}
The normalization used for the Fourier transform pair displayed in
expression (14) and (15) has been chosen to most directly reflect the
continuum relation of expression (11).

It follows from the convolution theorem that the spectral line shape,
$S^{res}$, for a particular set of dwell times $T_n$ is proportional
to a Fourier transform,

\begin{equation}
S_k^{res} = {1 \over T_{tot}} \sum_{n = -N/2 + 1}^{N/2} T_n exp(-i 2 \pi kn/N).
\end{equation}
With this normalization, the peak of the resolution function at $k=0$
is equal to unity. This resolution function plays the same role as the
transmission function $\tau(\nu)$ for the TF case. Just as for the TF
case, an effective width for the resolution function may be defined by
summing over all $k$ values,

\begin{equation}
\Delta\nu_{eff} = \delta \nu \sum_{k = -N/2 + 1}^{N/2} S_k^{res} =
{ T_0 \over T_{tot}} N \delta \nu.
\end{equation}
Although the case of uniform integration times is simplest for the FT
spectrometer, and indeed is the most common mode of operation of
laboratory FTIR instruments, it is not the most
efficient. Furthermore, for purposes of comparison with a TF
spectrometer, the resolution function (a sinc function) has negative
sidelobes, which cannot be realized by a physical transmission filter
function $\tau(\nu)$. There are many choices for the dwell time series
which produce non-negative spectral line shape functions which can be
physically realized as transmission filter profiles. One of the
simplest is the triangular apodization series, defined by

\begin{equation}
T_n = T_0 \left( 1 - {2|n| \over N} \right), \ n = -N/2+1,...,0,...N/2.
\end{equation}
The spectral line shape that results from this weighting is a
sinc-squared function.

\subsection{Fourier Transform Spectrometer Noise}

For a real interferogram, the discrete spectrum is hermitian, i.e.,
$Re(S_k) = Re(S_{-k})$.  While $Im (S_k) = - Im( S_{-k})$.  The point
$k=N/2$ corresponds to the Nyquist frequency.  For a perfectly
compensated beam splitter, with 100\% modulation efficiency and no
noise, the interferogram will also be symmetric. A real, symmetric
interferogram produces a real, symmetric spectrum. Noise in the
interferogram is real, and produces a hermitian contribution to the
calculated spectrum. Noise in the interferogram is not necessarily
symmetric, however, and thus contributes to both the real and the
imaginary parts of the calculated spectrum. By virtue of the linear
relation between interferogram and spectrum, and with the notation
that primed quantities represent noise contributions, the spectral
noise is simply the Fourier transform of the interferogram noise.

\begin{equation}
S'_k = {2 \over N} \sum_{n = -N/2 + 1}^{N/2} I'_n exp(-i 2 \pi kn/N).
\end{equation}

For a dual ported interferometer, with focal plane detectors having
equivalent noise performance characteristics, specifically having a
noise variance given by the sum of a read noise term, $n_r^2$, plus a
statistical noise term, the difference interferogram measurements have
the noise characteristics:

\begin{equation}
<I'_n> = 0. <I'_n I'_m> = \delta_{n,m}
\left( 2n_r^2 + {T_n\over T_0} I_0 \right).
\end{equation}
In the above expressions, the angle brackets represent an ensemble
average. It is assumed that the noise is uncorrelated for different
samples of the interferogram.  The statistical properties of the
spectral noise that follow from (19) and (20) are

\begin{equation}
<S'_k> = 0, < Re(S'_k)^2 > = \left( {2 \over N} \right)^2
\sum_{n = -N/2 + 1}^{N/2}
\left( 2n_r^2 + {T_n\over T_0} I_0 \right)
cos^2( 2\pi kn/N).
\end{equation}
Since for finite $k$ values, the $cos^2$ factor in expression (21)
oscillates much more rapidly as a function of $n$ than the factor
$T_n$, it may be well approximated by 1/2. With this approximation,
the spectral noise becomes independent of $k$, i.e., it is
``white''. The variance of the measured continuum spectrum thus is
given by

\begin{equation}
var( S(\nu_k)  ) =
{<  Re(S'_k)^2  > \over (T_0 \delta \nu)^2  } =
{ 2 \over ( \Delta \nu_{eff} T_{tot})^2}
\left(2Nn_r^2 + T_{tot} \int_0^\infty S(\nu) d\nu
\right).
\end{equation}
In the case that of a 1-sided interferogram, with samples

\begin{equation}
I_n = T_n I_\Delta (x_n), \ {\rm with\ } x_n = n \delta x,
\ n\ {\rm in\ the\ range\ } [0,N_1-1].
\end{equation}
The variance of the measured continuum spectrum is given by

\begin{equation}
var(S(\nu_k)) =
{ 1 \over ( \Delta \nu_{eff} T_{tot})^2}
\left(2N_1 n_r^2 + T_{tot} \int_0^\infty S(\nu) d\nu
\right).
\end{equation}
Although it may appear that the decrease in the variance has come
``for free'', there is really no greater information content, since
the density of independent spectral samples is only half as great in
the spectrum derived from the 1-sided interferogram. The difference in
the variance between 1-sided and 2-sided interferograms can be most
easily derived (for perfectly symmetrical interferograms) by averaging
each -n interferogram sample with the +n sample, and computing the
Fourier transform of the resulting 1-sided interferogram. The
statistical noise would be reduced by a factor of $1/\sqrt{2}$ for
each interferogram sample, and since only half as many readouts would
be required, the readout noise would be reduced a factor of two.

Expression (24), in the absence of read noise, matches expression (8)
obtained for the TF case. Expression (22), similarly matches
expression (7) for the TF case with a sampling interval $\Delta \nu_0
= {1 \over 2} \Delta \nu_{eff}$, as is appropriate for the more dense
sampling in frequency space. The interesting fact that the only
spectral line shape parameter that enters into the spectral noise for
a Fourier transform spectrometer is the effective width, $\Delta
\nu_{eff}$ , is novel, to this author's knowledge (e.g., Griffiths \&
de Haseth, 1986). The remarkable equivalence of the noise performance
over all of the various types of ideal imaging spectrometers may
perhaps be interpreted in terms of an ``information theory'' argument.

\section{Zodiacal Background and Detector Noise Terms}

\begin{figure}
\plotfiddle{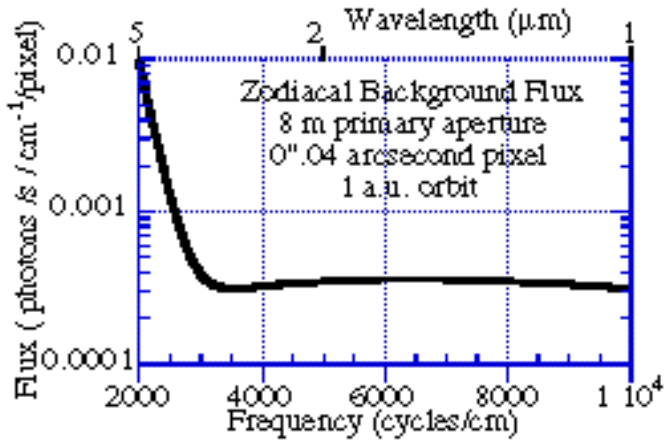}{2.5in}{0}{100}{100}{-150}{0}
\caption{The magnitude of the Zodiacal light background expected
for NGST in a 1AU orbit. }
\label{Bennett-fig2}
\end{figure}

The Zodiacal light produces a substantial limiting background flux for
NGST. For a 1 AU orbit, thermal emission from dust dominates at
wavelengths longer than about 3.5 $\mu$m, while for wavelengths shorter
than this, scattered sunlight produces the dominant background. An
estimate of this background spectrum is displayed in Figure
\ref{Bennett-fig2}.  The zodiacal background flux is constant, to good
approximation, over the range of frequencies from 3,000 to 10,000
cycles/cm, at a level of approximately $3x10^{-4}$ photon cm s$^{-1}$.
Detector noise performance levels anticipated for deployment on NGST
are displayed in the table below. The impact on the performance of the
various 3-d imaging systems generated by these background sources are
displayed in the next section.

{\small
\begin{table}
\begin{center}
\caption{NGST Detector Performance Expectations}
\medskip
\begin{tabular}{lll}
\hline
Case & Single Frame Read    & Dark Current \\
     & Noise, $n_1$ ($e^-$) & $I_d$ ($e^-$ s$^{-1}$) \\
\hline
Current & 15 & 0.1 \\
Goal    & 3  & 0.02 \\
\hline
\end{tabular}
\end{center}
\end{table}
}

\section{The Impact of Backgrounds on 3-d Spectrometers}

The signal to noise ratio, SNR, in the presence of backgrounds,
(including detector dark current, $I_d$, read noise, $n_r$, and zodiacal
background, $Z(\nu)$), for the TF and the DS case, is given by

\begin{equation}
SNR = {\eta\; QE(\nu)\; S(\nu) \Delta\nu_{eff} T_\nu
\over
\sqrt{ ((QE(\nu)(S(\nu) + Z(\nu))\Delta \nu_{eff})   +I_d)T_\nu + n_r^2    }
}.
\end{equation}
In this expression, the $QE(\nu)$ factor represents not only the quantum
efficiency of the detector elements themselves, but also includes all
other system transmission losses, such as non-unity reflectance for
reflecting surfaces, and non-zero absorption losses in any
transmitting elements. In general this factor may vary spectrally, and
in a fairly complex manner. Dispersive gratings, for example, will
have reflective losses for off-blaze angles of reflection. Tunable
filters have absorptive losses which tend to be greater for high
resolution broadly tunable filters. The efficiency factor $\eta$ in
expression (25) encapsulates all effects which lower the signal level
without concomitantly lowering the background level, such as surface
scattering losses, for example.

The signal to noise ratio in the presence of backgrounds for the FT
case is obtained from expression (24) by including the modulation
efficiency factor and QE effects, adding both the zodiacal background
and the detector dark current contributions to the total number of
photoelectrons observed, and including readout noise,

\begin{equation}
SNR = {\eta\; QE(\nu)\; S(\nu) \Delta\nu_{eff} T_{tot}
\over
\sqrt{   (\int QE(\nu)(S(\nu) + Z(\nu))\delta \nu   +2I_d)T_{tot} 
+ 2Nn_r^2}}.
\end{equation}
This expression reduces to the expression for the spectral SNR of
Graham, et al. (1998), under the simplifying assumptions made in that
article.  Fourier transform spectrometers have a modulation efficiency
factor which enters into the $\eta$ efficiency factor for the
spectrally resolved $SNR$, but not into the pan-chromatic $SNR$. The
magnitude of the read noise term which appears in expressions (25) and
(26) depends on the number of non-destructive readouts of the detector
array that are averaged to determine the estimated photo-current $I$
for a particular spectrometer setting. This number is a compromise
between integrating the photo-current for a longer time, which lowers
the variance, and averaging more readouts, which decreases the amount
of time available for integration. To good approximation, the optimum
number of readouts is determined by $I$, the single frame read noise,
$n_1$ and the time it takes to read out the array, $\Delta_t$, via
\begin{equation}
{\rm Optimal\ Number\ of\ Reads} = {n_1 \over \sqrt{I \Delta t  } }
\end{equation}
Therefore, $n_r^2 = n_1 \sqrt{I \Delta t}$.

Estimates of plausible realistic values for the system $QE$ and $\eta$
values are listed in Table 2 for operation in the $K$ band.  The NGST
main telescope transmission was calculated assuming that all of the
mirrors are gold coated. The dispersive spectrometer values are taken
from curves computed by Satyapal (1999), for the $K$ band at either
$R$=1000 or $R$=10,000, although they are somewhat optimistic compared
to the values routinely obtained at large ground based
telescopes.\footnote{See, for example, the grating efficiency curves measured
for the 2dF gratings, at the AAO website location:
http://www.aao.gov.au/local/www/ras/gratings/gratings.html.}
The tunable filter efficiencies are estimated on the
basis of Northrop Grumman tunable Fabry-Perot performance values (Madonna \& Ryan). The
somewhat surprisingly low QE for the tunable filter may be attributed
to the fact that the effective number of surfaces seen by the
transmitted light is approximately equal to the finesse, and thus
extremely high quality surfaces are necessary to produce low losses at
high finesse. The FT modulation efficiency corresponds to that of a
30$^\circ$ incident angle NIR-MidIR CsI Bomem beamsplitter (Villemaire 1999).

{\small
\begin{table}
\begin{center}
\caption{Spectrometer Efficiency Assumptions ($K$ band)}
\medskip
\begin{tabular}{lll}
\hline
Case & QE    & $\eta$ \\
     & detector+optics &  \\
\hline
TF & 0.35 & 1 \\
DS & 0.6  & 0.7 \\
FT & 0.7  & 0.95 \\
\hline
\end{tabular}
\end{center}
\end{table}
}

\begin{figure}
\plotfiddle{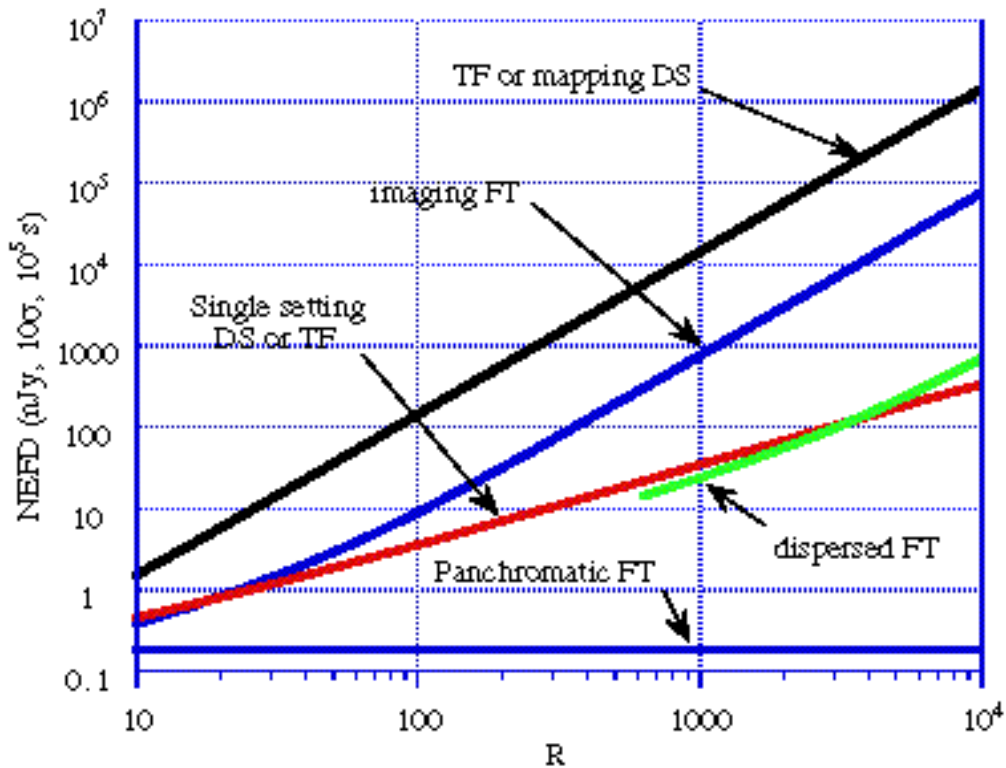}{3.5in}{0}{100}{100}{-250}{0}
\caption{ The Noise Equivalent Flux Density for a variety of 3d
imaging approaches are displayed.  The detector performance parameters
are: 15 e$^-$ read noise per readout, 0.1 e$^-$ s$^{-1}$ dark current.
All observations extend over the $K$ band, centered at 2.2 $\mu$m, and
assume a total observation time of $10^5$ s. The time per readout is
assumed to be 1 s. The QE and efficiency, $\eta$, values are those of
Table 2 for each of the systems. TF = tunable filter, DS=dispersive
spectrometer, FT=Fourier transform spectrometer. The ``mapping DS''
corresponds to scanning over a number of field of view settings equal
to the number of spectral channels for the TF at each resolution, and
thus has an NEFD equal to that of the TF spectrometer. It is only
through this factor that the array size enters, as was shown in figure
1. The single setting TF corresponds to the observation of a full
field of view, but through only a single filter. The panchromatic FT
line indicates the NEFD sensitivity for the full K band imaging that
is obtained as a function of the resolution of the imaging FT
system. This sensitivity is almost independent of the resolution,
since the loss of exposure time involved in the multiple readouts for
the spectroscopy is a small fraction of the total observation
time. The dispersed FT curve corresponds to the addition of an $R$ =
630 prism in the collimated space of the imaging FT optical train, and
the addition of a slit at an image plane. The dispersed FT case with
slit, would not produce the panchromatic FT full field imaging. A
dispersed FT operated ala a slit-less objective prism would produce
the panchromatic FT full field imaging.  }
\label{Bennett-fig3}
\end{figure}

The Noise Equivalent Flux Density, $NEFD$, at a particular significance
level is derived from the $SNR$ equations by solving for the flux $S$
which produces the given significance level. The NEFD for observations
in the K band at 2.2 $\mu$m, (as one example) at the 10 $\sigma$ level,
for a variety of imaging spectrometer options are displayed in Figure
\ref{Bennett-fig3} as a function of spectral resolution. From these
curves, for a particular problem of interest, it is easy to select the
optimum instrumental configuration.

At the lowest spectral resolution At the lowest spectral resolution,
all of the 3-d instruments converge to the performance of an $R$=5,
$K$ band camera. At the highest spectral resolution, the DS has the
best performance for spectroscopy, although only for the small number
of objects that may be contained ``within the slit''. This fact is the
basis for the current pre-eminence of Multi-Object Spectrometers and
Integral Field Units in high resolution astronomical spectroscopy. For
the purpose of imaging in a very narrow, single emission line band,
the TF provides an $NEFD$ performance equivalent to that of the DS, but
for every pixel in the field of view. For the purpose of obtaining
complete spectra for every pixel in the field of view, the FT
instrument substantially outperforms the TF or the ``mapping DS'',
(whose performance becomes essentially equivalent to the TF). The
point of equivalence between the imaging FT and the DS comes at the
point for which the number of settings of the DS is equal to the
square of the ratio in performance between the single setting DS and
the imaging FT.

For any resolution, the imaging FT instrument has the
advantage that not only are spectra obtained for every pixel in the
field of view, but that very deep K-band imaging (in this example, but
it could be $J$, $H$, $L$, etc. or the entire 0.6-5.5 $\mu$m range) is
simultaneously acquired. In many of the design reference missions for
NGST, the data for both deep imaging and spectroscopy may be acquired
simultaneously. The fact that such imaging is produced for every
resolution setting of the FT instrument is indicated in Figure
\ref{Bennett-fig3} by the lowest $NEFD$ curve labeled ``panchromatic FT''.

At the highest spectral resolution, the relatively strong signals
required imply that for many fields of interest to NGST, the angular
density of observable objects will be small enough that at most one
object is expected per field of view. In this situation, it is not
helpful to obtain spectra for every pixel in the field of view, and
the spatial multiplexing of the imaging FT is not useful.

A very interesting hybrid approach (e.g., Beer 1992) is possible,
however, which takes advantage of the best features of all of the 3-d
imaging approaches. This is the combination of an objective prism with
an imaging FT spectrometer. A relatively modest dispersion across one
dimension of the image plane serves to reduce the spectral bandpass
acceptance that is involved in the noise term for the FT spectrometer.
With a slit at an image plane, the ``panchromatic'' output of the FT
spectrometer would yield the same results as an ordinary prism
spectrometer, while the Fourier transformed interferograms would
enable much higher spectral resolution at much reduced NEFD. The curve
labeled ``dispersed FT'' in Figure \ref{Bennett-fig3} corresponds to
the assumption that a prism of dispersion equal to that of CaF$_2$ is
placed in the collimated space of an imaging FT, and that the slit
width is equal to one pixel. For objects which have much higher
intensity than their surroundings, slit-less objective prism style
measurements are also possible.

There are slight displacements of the curves for the various
spectrometer, and imaging spectrometer configurations, depending on
the choices for the detector performance parameters, and system
efficiency values. Using the NGST ``goal'' detector performance
parameters instead of the ``current'' performance values slightly
improves the sensitivity of the TF, the DS, and the dispersed FT, but
produce very little change in the imaging FT case. On the other hand,
using grating efficiencies closer to those typical of ground based
telescopes, lowers the DS curve, but not the TF, the imaging FT, or
the dispersed FT curves. The mapping DS curve does not take into
account any in-efficiencies with the precision re-pointing between
observations.

In conclusion, the ability of a single instrument concept, composed of
a filter wheel, programmable slit, dispersive prism, and Michelson
interferometer, to deliver the performance of a wide field camera, the
performance of a moderate resolution, full field imaging spectrometer,
and the performance of a high resolution, limited field spectrometer
seems to make this choice nearly obligatory for NGST.

This work was performed under the auspices of the U.S. Department of
Energy under Contract No W-7405-Eng-48. I thank my IFIRS colleagues
for many stimulating discussions and astronomical tutoring: J. R. Graham,
M. Abrams, J. Carr, K. Cook, A. Dey, R. Hertel, N. Macoy, S. Morris,
J. Najita, A. Villemaire, E. Wishnow, and R.Wurtz. I also thank
J. Mather for the provocative suggestion to consider the dispersed FT
option.

\end{document}